\documentclass[aps,prl,groupedaddress,draft]{revtex4}
\begin{document}

%\hoffset-1cm

% Yale printer values
\voffset1.5cm
%\draft
%\preprint{nucl-th/99????}
\title{Eikonal but not: a complementary view of high energy evolution}
\author{Tolga Altinoluk, Alex Kovner and Javier Peressutti}
\affiliation{ Physics Department, University of Connecticut, 2152 Hillside road, Storrs, CT 06269, USA}

\date{\today}
\begin{abstract}
The high energy evolution equations that describe the evolution of hadronic amplitudes with energy are derived assuming eikonal interaction of the evolved hadronic wave function with the target. In this note we remark that this derivation allows a different interpretation, whereby the hadronic wave function is not evolved, but instead the evolution acts on the $S$ - matrix operator. In this approach, analogous to the Heisenberg picture of Quantum mechanics, the scattering is not eikonal and additional boost provides for radiation of more gluons in the final state.
\end{abstract}
\maketitle

%%%%%%%%%%%%%%%%%%%%%%%%%%%%%%%%%%%%%%%%%%%%%%%%%%%%%%%%%%%%%%%%%%%%%%
This short note continues the program of investigating high energy evolution of hadronic scattering observables\cite{gribov}, \cite{BFKL},\cite{GLR},\cite{mv}, \cite{Bartels},\cite{BKP},\cite{balitsky}, \cite{kovchegov},\cite{JIMWLK},\cite{cgc}. 
In the past several years the activity in this  area concentrated on attempting to include the effects of Pomeron loops in the evolution with many contributions by several groups \cite{shoshi},\cite{pomloops},\cite{stat},\cite{klwmij},\cite{duality},\cite{remarks},\cite{remarks2},\cite{genya}.
In particular a recent paper \cite{foam} gives a derivation of the evolution of a hadronic light cone wave function valid for all parametrically interesting values of the valence color charge density, from the weak charge density (KLWMIJ) limit  to the large charge density (JIMWLK) limit.
How to use this information to derive the evolution of scattering amplitude is still an open question. 
The physical picture of the evolution as described for example in \cite{zakopane}, is that the increase in energy is used to boost the projectile wave function. The boost leads to increase of the number of gluons in the projectile wave function. These gluons then scatter eikonally when propagating through the color field of the target.

The statement that the high energy scattering in non abelian theories is eikonal frequently causes eyebrows to lift, as it is well known from diagrammatic calculations that at high energy logarithmic contributions to the $S$ matrix come from final states with extra gluons. The confusion here stems from the fact that even though each parton scatters eikonally, the whole projectile state of course does not and so inelastic final states certainly contribute to the scattering.
The evolved wave function contains more partons. Those partons decohere while propagating through the target field, even though individual partons do not emit gluons. Due to this decoherence the majority of final states are inelastic. We point out in this note that the derivations of \cite {JIMWLK} and \cite{klwmij}
can be also given another interpretation. This is akin to the Heisenberg picture of quantum mechanics. Here it is not the projectile state that is evolved, but rather the $S$-matrix operator. The $S$ matrix thus has explicitly noneikonal contributions which are responsible for emission of extra gluons into the final state.

Consider calculation of the expectation value of any observable in the wave function of the incoming probe $|v\rangle$. We will be interested only in observables which depend on the color charge density operator $j^a(x)$ and not on any other property of the wave function
\begin{equation}
\bar O=\langle v\vert O[j]\vert v\rangle
\end{equation}
An example of such an observable is the eikonal $S$ - matrix in the external field $\alpha$
\begin{equation}\label{eikonal}
S=\exp\bigg\{i\int_x j^a(x)\alpha^a(x)\bigg\}
\end{equation}
In the approach of \cite{JIMWLK} the calculation of the scattering amplitude for a hadron-hadron scattering is given by
\begin{equation}
{\cal S}=\langle \bar S\rangle_{\alpha}
\end{equation}
where the weight for the averaging over the fields $\alpha$ is furnished by the target wave function.

The projectile wave function at certain rapidity contains partons with longitudinal momenta larger than a certain cutoff $\Lambda$.
As discussed many times in the literature, the standard way to evolve the observable with rapidity is to boost the projectile wave function. Under boost, the longitudinal momenta of the existing partons are scaled by the Lorentz $\gamma$ factor, while the newly opened phase space is filled by soft gluons. 
\begin{equation}
\vert v\rangle\rightarrow \Omega_Y[j, a^\dagger, a]\vert v\rangle
\end{equation}
The operator $\Omega_Y$ creates the soft gluons in the projectile wave function. It is a Bogoliubov type operator and has been calculated in \cite{foam} for arbitrary charge density $j$. In the low density limit, that is when $j=O(g)$ the form of this operator has been found in \cite{klwmij}. In this case it reduces to the coherent operator 
\begin{equation}
\Omega_Y\rightarrow 
C_Y=\exp \left\{2i\int d^2xb_i^a(x)\int{dk^+\over  2\pi}{1\over |2k^+|^{1/2}} [a^{\dagger a}_i(k^+, x)+a^a_i(k^+, x)]\right\}
\end{equation}
where the integration is over the phase space opened by boost to rapidity $Y$,
and the "classical field" $b$ is determined by the color charge density through
\begin{equation}
b_i^a(x)={1\over 2\pi}\int d^2y {x_i-y_i\over (x-y)^2}j^a(y), \ \
\end{equation}
and the soft gluon creation and annihilation operators satisfy
\begin{equation}
\left[ a_a^i(k^+, x), a_b^{j\, \dagger}(p^+, y)
   \right] = (2\pi)\, \delta_{ab}\, \delta^{ij}\, \delta(k^+ -p^+)\,
             \delta^{(2)}(x - y)\, .
\end{equation}
The observable $O$ calculated in the state boosted to rapidity $Y$ is given by
\begin{equation}\label{evolvO}
\bar O_Y=\langle v\vert \Omega^\dagger_Y O \Big[ j^a(x)+ g\! \int {dk^+\over 2\pi}a^\dagger(k^+,x) T^a a(k^+,x) \Big] \Omega_Y \vert v\rangle
\end{equation}
with $T^a_{bc}=if^{abc}$.
The charge density operator $j$ is shifted here since the newly produced gluons also carry color.

Since we are interested in the dependence of observables on rapidity, the rapidity in this approach plays the role of time - the parameter of the evolution.
The operator $\Omega$ then plays the role of the evolution operator in rapidity. The state 
$\vert v\rangle_Y=\Omega_Y\vert v\rangle$ is analogous to the time dependent state in the Schroedinger representation of Quantum mechanics. Thus the picture of the evolution of the scattering amplitude for example, is that the projectile state is evolved while the scattering matrix operator always stays eikonal. There are more partons (gluons) in the evolved state, and the scattering matrix element changes with rapidity, even though the partons scatter eikonally and do not emit any extra gluons during the propagation through the target.

One can give an alternative interpretation to eq.(\ref{evolvO}). In particular we can think of the operator $\Omega_Y$ as acting on the observable $O$ rather than on the state $\vert v\rangle$. In this interpretation it is the operator $O$ that is evolved to higher rapidity rather than the state $\vert v\rangle$. This is analogous to the Heisenberg picture of Quantum mechanics. Thus, for example if at low enough rapidity we take the operator $O$ as eikonal scattering matrix eq.(\ref{eikonal}) its evolution with $\Omega$ produces a new $S$ matrix operator which is not eikonal anymore, but rather allows for emission of final state gluons. This is straightforward to see using the explicit form of the action of the operator $\Omega$ on the fields which has been given in \cite{foam}. For simplicity instead of general $\Omega$ we will use here its low density version $C$. Thus to order $g^2$ we have
\begin{eqnarray}\label{evolvj}
C^\dagger \bigg[ j^a(x)+g\!\int {dk^+\over 2\pi} a^\dagger(k^+,x) T^aa(k^+,x) \bigg]C &=&  
				j^a(x)+g\!\int {dk^+\over 2\pi} a^\dagger(k^+,x) T^aa(k^+,x)\nonumber \\
&& \!\!\!\!\!\!\!\!\!\!\!\!\!\!\!\!\!\!    -2i\,g\, b^b_i(x)T^a_{bc}A^c_i(x,0)-2i
{g\over 2\pi}\!\int d^2y{y_i-x_i\over (y-x)^2}j^b(x)T^a_{bc}A^c_i(y,0)]\nonumber\\
&& \!\!\!\!\!\!\!\!\!\!\!\!\!\!\!\!\!\!    -{g^2\over 2\pi^2}\!\left[\int_{z,y}{(y-z)_i(x-z)_j\over (y-z)^2(x-z)^2}A^j_c(x,0)T^a_{cb}T^d_{be}A^e_i(y,0)j^d(z)+(x\rightarrow z)\right]
\end{eqnarray}
where we have defined
\begin{equation}
A^c_i(x,0)=\int {dk^+\over 2\pi}{1\over \sqrt {2k^+}}[a_i^c(k^+,x)+a_i^{\dagger c}(k^+,x)]
\end{equation}
Obviously the "rapidity evolved" color charge density contains terms linear and quadratic in the soft gluon creation operator. Thus the "rapidity evolved" $S$-matrix 
\begin{equation}
S_Y=C^\dagger SC=\exp\bigg\{i\int_x C^\dagger j^a (x)C\alpha^a(x)\bigg\}
\end{equation}
describes not only eikonal propagation of original partons, but also emission of soft gluons 
by these partons while they propagate through the target field. As is obvious from eq.(\ref{evolvj}) the amplitude of such emission is proportional to the charge density in the projectile, the field strength in the target and the rapidity interval $Y$ over which the process has been evolved.

We note that the original derivation of the JIMWLK evolution equation that describes the high density limit was in fact performed by calculating the change of the color charge density due to the evolution. In this sense the original derivation uses the Heisenberg picture of the evolution.

The derivation of the KLWMIJ equation on the other hand was given in the Schroedinger picture. In the rest of this note we will rederive this equation in the Heisenberg picture. The purpose of this exercise is twofold. Firstly, we want to put both derivation in the same framework. Secondly and more importantly we anticipate that the generalization of the high energy evolution to the general case (beyond JIMWLK and KLWMIJ limits) will be more conveniently done in the Heisenberg picture. The reason is that in \cite{foam} the action of the operator $\Omega$ on all the degrees of freedom of the theory has been calculated. Thus we know the explicit form of the evolved color charge density. On the other hand the explicit form of the operator $\Omega$ in terms of dynamical fields has not been found, and therefore its explicit action on the wave function $\vert v\rangle$ is not available. Thus we want to make our hands a little dirty  and gain some more experience working with the Heisenberg representation.

Any observable $O[j]$ can be expanded in Taylor series in powers of $j$. Thus it is enough to consider 
\begin{equation}
O_n=j^{a_1}(x_1)...j^{a_n}(x_n)
\end{equation}
When evolved to small rapidity $\Delta Y$
\begin{equation}
O^n\rightarrow O_{\Delta Y}^n=C^\dagger J^{a_1}(x_1)...J^{a_n}(x_n)C
\end{equation}
with 
\begin{equation}
J^a(x)=j^a(x)+g\!\int {dk^+\over 2\pi} a^\dagger(k^+,x) T^aa(k^+,x)
\end{equation}

Our aim is to represent the expectation value of the evolved operator $O^n$ in the valence state in terms of a "Hamiltonian" acting on the expectation value of the unevolved operator
\begin{equation}\label{ev}
\langle v\vert O^n_{\Delta Y}\vert v \rangle =\langle v\vert O^n\vert v \rangle 
+\Delta Y\langle v\vert H\Big[j,{\delta\over\delta j}\Big]O^n\vert v \rangle 
\end{equation}
The above equation is somewhat little cryptic, since $j$ is a quantum operator and differentiation with respect to quantum operator has to be defined. To make its meaning precise we remind the reader that averaging over the valence wave function can be done in the path integral representation\cite{klwmij}. For this purpose one introduces an ordering variable $t$ and 
considers classical functions $j^a(x,t)$. The average is then written as
\begin{equation}\label{funin}
\langle v\vert j^{a_1}(x_1)...j^{a_n}(x_n) \rangle=\int Dj W[j(x,t)] j^{a_1}(x_1,t_1)...j^{a_n}(x_n,t_n)
\end{equation}
such that $0<t_1<t_2<...<t_n<1$. The values of the variable $t$ are ordered in the same way as the position of the appropriate charge density operator in the operator product. This is necessary to account for noncommutativity of the quantum charge density operators. The weight function $W$ is determined by the valence wave function. It is not arbitrary but has to satisfy particular constraints. This has been discussed at length in \cite{remarks2}. For our purposes it is only important to know that such a representation exists. The exact values of the time variable $t_i$ are also unimportant as long as the ordering is preserved. Given this the evolution eq.(\ref{ev}) takes the form of a functional differential equation
\begin{equation}\label{evw}
{\delta\over\delta Y}W[j]=H\Big[ j,{\delta\over\delta j} \Big] W[j]
\end{equation}
In eq.(\ref{evw}) $j$ is a classical function which depends on the transverse coordinates $x$ as well as the ordering coordinate $t$ and functional differentiation is now well defined.

To order $g^2$ we have
\begin{eqnarray}\label{evolvedj}
&& \!\!\!\!\!\!\!\!  \langle v\vert\, C^\dagger J^{a_1}(x_1)...J^{a_n}(x_n)C\,\vert v\rangle=\\
&&\sum_{m<k,m=1}^{n-1}\langle v \vert j^{a_1}(x_1)...j^{a_{m-1}}(x_{m-1})  \bigg[ 2i\int_xb^i_a(x)A^a_i(x,0), J^{a_m}(x_m)  \bigg]J^{a_{m+1}}(x_{m+1})...J^{a_{k-1}}(x_{k-1})\times\nonumber\\
&&\times  \bigg[ 2i\int_xb^i_a(x)A^a_i(x,0), J^{a_k}(x_k)  \bigg] j^{a_{k+1}}(x_{k+1})...j^{a_n}(x_n)\vert v\rangle\nonumber\\
&&+{1\over 2}\sum_{m=1}^n \langle v \vert j^{a_1}(x_1)...j^{a_{m-1}}(x_{m-1})  \bigg[ 2i\int_xb^i_a(x)A^a_i(x,0), \bigg[ 2i\int_yb^j_b(y)A^b_j(y,0), J^{a_m}(x_m) \bigg] \bigg] j^{a_{m+1}}(x_{m+1})...j^{a_n}(x_n)\vert v\rangle\nonumber
\end{eqnarray}
Here the charge density operators acting on the state $\vert v\rangle$ either to the right or to the left contain only valence part since the soft gluon contribution to charge density vanishes in the state $\vert v\rangle$ which contains no soft gluons.
Let us first concentrate on the double sum term in eq.(\ref{evolvedj}). First, we have to average over the soft gluon Hilbert space. This is quite straightforward. Only the term involving the annihilation operator $a$ contributes in $A(x,0)$ in the first commutator, and only the term with $a^\dagger$ contributes in the second commutator. The soft charge density operators between these two insertions pick up the contribution from the one gluon state created from $\vert v\rangle$ by the $a^\dagger$ term, and thus are simply shifted by the one gluon charge. Once we have averaged over the soft gluons, the rest of the average can be represented as a functional integral of the form eq.(\ref{funin}). We can thus write 
\begin{eqnarray}
&&\qquad\qquad -{1\over \pi}Y\int Dj W[j]\sum_{m<k,m=1}^{n-1}\Bigg\{  \nonumber\\
&&\int_xb^i_a(x,t_m-\epsilon)R^{ab}(x,t_m-\epsilon,t_k-\epsilon)b^i_b(x,t_k-\epsilon)
-\int_xb^i_a(x,t_m+\epsilon)R^{ab}(x,t_m+\epsilon,t_k-\epsilon)b^i_b(x,t_k-\epsilon)     \nonumber\\
&-&
\int_xb^i_a(x,t_m-\epsilon)R^{ab}(x,t_m-\epsilon,t_k+\epsilon)b^i_b(x,t_k+\epsilon)
+\int_xb^i_a(x,t_m+\epsilon)R^{ab}(x,t_m+\epsilon,t_k+\epsilon)b^i_b(x,t_k+\epsilon)
\Bigg\}\times\nonumber\\
&&\qquad \times j^{a_1}(x_1,t_1)...j^{a_{m-1}}(x_{m-1},t_{m-1}) j^{a_m}(x_m,t_m)...j^{a_{k}}(x_{k-1},t_{k-1})j^{a_{k}}(x_{k},t_k)...j^{a_{n}}(x_n,t_n)
\end{eqnarray}
In this expression the time variables are strictly ordered $t_1<t_2...<t_n$. 
The operator 
\begin{equation}
R(x,t,t')={\cal P}\exp\bigg\{ g\! \int_t^{t'} d\tau{\delta\over\delta j^a(x,\tau)}T^a\bigg\}
\end{equation}
shifts the color charge density by the charge of a single gluon. The time coordinate of the field $b$ is shifted by an infinitesimal amount $\epsilon$ with respect to the coordinate of the nearest charge density so that the ordering of time coordinates follows the operator ordering of eq.(\ref{evolvedj}). The differences between two pairs of terms combine into time derivatives if the summations are changed to time integrals so that we can write:
\begin{equation}\label{doubleder}
-Y{1\over 2\pi}\int Dj W[j]\int_0^1 dt\int_0^1 dt'{\partial\over \partial t}{\partial\over \partial t'}\int_xb_i^a(x,t)R^{ab}(x,t,t')b^b_i(x,t')j^{a_1}(x_1,t_1)...j^{a_{n}}(x_{n},t_{n})
\end{equation}

Although we have explicitly considered only the first term in eq.(\ref{evolvedj}), it is straightforward to see that the second (double commutator) term reproduces the $t=t'$ contribution in eq.(\ref{doubleder}).
 Integration over $t$ and $t'$ is straightforward with the result
 \begin{equation}
 H\Big[j,{\delta\over \delta j}\Big]={1\over 2\pi}\bigg[ 2\int_x b^a_i(x,0)R^{ab}(x,0,1)b^b_i(x,1)-b^a_i(x,0)b^a_i(x,0)-b^a_i(x,1)b^a_i(x,1)\bigg]
 \end{equation}
 This is precisely the KLWMIJ Hamiltonian first derived in \cite{klwmij}.
 
 This concludes the derivation of the KLWMIJ Hamiltonian in the Heisenberg representation of the evolution. We hope that this exercise will be useful for future derivation of the general evolution Hamiltonian including the Pomeron loop contributions using the results of \cite{foam}.

%%%%%%%%%%%%%%%%%%%%%%%%%%%%%%%%%%%%%%%%%%%%%%%%%%%

%

\begin{thebibliography}{99}

\bibitem{gribov}   V.~N.~Gribov,
  %``A Reggeon Diagram Technique,''
  Sov.\ Phys.\ JETP {\bf 26}, 414 (1968)
  [Zh.\ Eksp.\ Teor.\ Fiz.\  {\bf 53}, 654 (1967)].
  %%CITATION = SPHJA,26,414;%%
  
\bibitem{BFKL}
 E. A. Kuraev, L. N. Lipatov, and F. S. Fadin,  Sov. Phys. JETP
                {\bf 45} (1977) 199 ; \\
Ya. Ya. Balitsky and L. N. Lipatov,
               {  Sov. J. Nucl. Phys.}\, {\bf 28} (1978) 22;
                 L.~N.~Lipatov
  %``Reggeization Of The Vector Meson And The Vacuum Singularity In Nonabelian
  %Gauge Theories,''
  Sov.\ J.\ Nucl.\ Phys.\  {\bf 23}, 338 (1976)
  [Yad.\ Fiz.\  {\bf 23}, 642 (1976)].
  %%CITATION = SJNCA,23,338;%%
  L.~L.~Frankfurt and V.~E.~Sherman,
  %``Reggeization Of Vector Meson And Vacuum Singularity In Renormalizable
  %Yang-Mills Models,''
  Sov.\ J.\ Nucl.\ Phys.\  {\bf 23} (1976) 581.
  %%CITATION = SJNCA,23,581;%%
  \bibitem{GLR}  L.V.~Gribov, E.~Levin and M.~Ryskin, Phys. Rep.
100:1,1983; A. H. Mueller and J. Qiu,  Nucl. Phys. {\bf B 268} (1986) 427.


\bibitem{mv} L.~McLerran and R.~Venugopalan, Phys. Rev. D49:2233-2241,1994 
% e-Print Archive: hep-ph/9309289
; Phys.Rev.D49:3352-3355,1994. 

\bibitem{Bartels} J. Bartels, 
  %``High-Energy Behavior In A Nonabelian Gauge Theory. 1. T (N--->M) In The
  %Leading Log Normal S Approximation,''
  Nucl.\ Phys.\ B {\bf 151}, 293 (1979);
J.~Bartels, Z.Phys. C60, 471, 1993; J.~Bartels and M.~Wusthoff, Z. Phys. {\bf C 66}, 157, 1995;  J.~Bartels and C.~Ewerz, JHEP  9909, 026, 1999
[e-Print Archive:hep-ph/9908454].

  %%CITATION = NUPHA,B151,293;%%


\bibitem{BKP}
J.~Bartels,
% ``High-Energy Behavior In A Nonabelian Gauge Theory. 2. First 
%Corrections To
%T(N--->M) Beyond The Leading Lns Approximation,''
%
Nucl.\ Phys.\  {\bf B175}, 365 (1980);\,\,\,\,
%%CITATION = NUPHA,B175,365;%%
J.~Kwiecinski and M.~Praszalowicz,
% ``Three Gluon Integral Equation And Odd C Singlet Regge Singularities In
%QCD,''
%
Phys.\ Lett.\  {\bf B94}, 413 (1980);
%%CITATION = PHLTA,B94,413;%%
 L.~N.~Lipatov,
  %``Asymptotic behavior of multicolor QCD at high energies in connection with
  %exactly solvable spin models,''
  JETP Lett.\  {\bf 59}, 596 (1994)
  [Pisma Zh.\ Eksp.\ Teor.\ Fiz.\  {\bf 59}, 571 (1994)]; L.~D.~Faddeev and G.~P.~Korchemsky,
  %``High-energy QCD as a completely integrable model,''
  Phys.\ Lett.\ B {\bf 342}, 311 (1995)
  [arXiv:hep-th/9404173];  G.~P.~Korchemsky,
  %``Quasiclassical QCD pomeron,''
  Nucl.\ Phys.\ B {\bf 462}, 333 (1996)
  [arXiv:hep-th/9508025];
  %%CITATION = HEP-TH 9508025;%%
  %``Bethe ansatz for QCD pomeron,''
  Nucl.\ Phys.\ B {\bf 443}, 255 (1995)
  [arXiv:hep-ph/9501232];
  %%CITATION = HEP-PH 9501232;%%
  %``Integrable structures and duality in high-energy {QCD},''
  Nucl.\ Phys.\ B {\bf 498}, 68 (1997)
  [arXiv:hep-th/9609123];
  %%CITATION = HEP-TH 9609123;%%
G.~P.~Korchemsky, J.~Kotanski and A.~N.~Manashov,
%``Compound states of reggeized gluons in multi-colour QCD as ground  states 
%of noncompact Heisenberg magnet,''
Phys.\ Rev.\ Lett.\  {\bf 88} (2002) 122002
[arXiv:hep-ph/0111185]; H.J. de Vega and L.N. Lipatov, Phys.Rev.D64:114019,2001, [arXive: hep-ph/0107225]; Phys.Rev.D66:074013,2002.
[arXive: hep-ph/0204245]
  
  




  
 
 
 

%\bibitem{BLW}
%  J.~Bartels, L.~N.~Lipatov and M.~Wusthoff,
  %``Conformal Invariance of the Transition Vertex $2 \to 4$ gluons,''
%  Nucl.\ Phys.\ B {\bf 464}, 298 (1996)
%  [arXiv:hep-ph/9509303].
  %%CITATION = HEP-PH 9509303;%%


%\bibitem{BV}
%  M.~A.~Braun and G.~P.~Vacca,
  %``Triple pomeron vertex in the limit N(c) $\to$ infinity,''
%  Eur.\ Phys.\ J.\ C {\bf 6}, 147 (1999)
%  [arXiv:hep-ph/9711486].
  %%CITATION = HEP-PH 9711486;%%


%\bibitem{BBV}
%  J.~Bartels, M.~Braun and G.~P.~Vacca,
  %``Pomeron vertices in perturbative QCD in diffractive scattering,''
%  Eur.\ Phys.\ J.\ C {\bf 40}, 419 (2005)
%  [arXiv:hep-ph/0412218].
  %%CITATION = HEP-PH 0412218;%%

%\bibitem{BLV} J.~Bartels, L.~N.~Lipatov and G.~P.~Vacca,
  %``Interactions of Reggeized gluons in the Moebius representation,''
%  Nucl.\ Phys.\ B {\bf 706}, 391 (2005)
%  [arXiv:hep-ph/0404110].
  %%CITATION = HEP-PH 0404110;%%

%\bibitem{dipoles}  A. Mueller, {\it 
%Nucl. Phys.} {\bf B335} 115 (1990); {\it ibid} {\bf B 415}; 
%373 (1994); {\it ibid} {\bf B437} 107 (1995).


\bibitem{balitsky} I. Balitsky, {\it Nucl. Phys.}  {\bf B463} 99 (1996); 
% e-Print Archive: hep-ph/9509348; in
{\it Phys. Rev. Lett.} {\bf 81} 2024 (1998); % e-Print Archive: hep-ph/9807434;in
{\it Phys. Rev.}{\bf D60} 014020 (1999).


\bibitem{kovchegov}
  Y.~V.~Kovchegov,
  %``Unitarization of the BFKL pomeron on a nucleus,''
  Phys.\ Rev.\ D {\bf 61}, 074018 (2000)
  [arXiv:hep-ph/9905214].
  %%CITATION = HEP-PH 9905214;%%

\bibitem{JIMWLK} J. Jalilian Marian, A. Kovner, A.Leonidov and H.
Weigert,
{\it Nucl. Phys.}{\bf  B504} 415 (1997); % e-Print Archive: hep-ph/9701284
{\it Phys. Rev.} {\bf D59} 014014 (1999); % e-Print Archive: hep-ph/9706377
J. Jalilian Marian, A. Kovner and H. Weigert, {\it Phys. Rev.}{\bf D59} 
014015 (1999); 
% e-Print Archive: hep-ph/9709432;
A. Kovner and J.G. Milhano, {\it Phys. Rev.} {\bf D61} 014012 (2000) .
% e-Print Archive: hep-ph/9904420.
 A. Kovner, J.G. Milhano and H. Weigert,
{\it Phys.Rev.} {\bf D62} 114005 (2000); 
 H. Weigert, {\it Nucl.Phys.} {\bf A 703} (2002) 823.
 

\bibitem{cgc}  E.Iancu, A. Leonidov and L. McLerran, {\it Nucl. Phys.} 
{\bf A 692} (2001) 583; {Phys. Lett.} {\bf B
510} (2001) 133;
E. Ferreiro, E. Iancu, A. Leonidov, L. McLerran;  
{\it Nucl. Phys.}{\bf A703} (2002) 489.

%\bibitem{yinyang} A. Kovner and M. Lublinsky, e-Print Archive: hep-ph/0512316 ;  Nucl.Phys.A779:220-243,2006, e-Print Archive: hep-ph/0604085

\bibitem{shoshi} A. Mueller and A.I. Shoshi, Nucl. Phys. B {\bf 692}, 175 (2004).
 
\bibitem{pomloops}A. Mueller, A. Shoshi and S. Wong, Nucl.\ Phys.\ B {\bf 715}, 440 (2005);
 E. Levin and M. Lublinsky; Nucl. Phys. A {\bf 763}, 172 (2005) 
[arxive hep-ph/0501173]; E. Iancu and  D. N. Triantafyllopoulos,  Phys.\ Lett.\ B {\bf 610}, 253 (2005); 
Nucl.\ Phys.\ A {\bf 756}, 419 (2005); E.~Iancu, G.~Soyez and D.~N.~Triantafyllopoulos,
  %``On the probabilistic interpretation of the evolution equations with Pomeron
  %loops in QCD,'' 
  arXiv:hep-ph/0510094;
  
\bibitem{stat} E. Iancu, A.H. Mueller and S. Munier, Phys.\ Lett.\ B {\bf 606}, 342 (2005);
E. Brunet, B. Derrida , A.H. Mueller and S. Munier;  Phys.Rev.E73:056126,2006.
e-Print Archive: cond-mat/0512021
  
\bibitem{klwmij}
A. Kovner and M. Lublinsky; Phys.Rev.D71:085004,2005.
e-Print Archive: hep-ph/0501198

\bibitem{duality} A. Kovner and M. Lublinsky; Phys. Rev. Lett.{\bf 94}, 181603 (2005)

\bibitem{remarks}  A. Kovner and M. Lublinsky; JHEP 0503:001,2005.
e-Print Archive: hep-ph/0502071


\bibitem{remarks2} A. Kovner and M. Lublinsky; Nucl.Phys.A767:171-188, 2006.

\bibitem{genya}  E. Levin;  Nucl.Phys. A763 (2005) 140-171 ; e-Print: arXiv:hep-ph/0502243;
     M. Kozlov, E. Levin (Tel Aviv U.)
     Nucl.Phys. A779 (2006) 142-176 ; e-Print: arXiv:hep-ph/0604039 
M. Kozlov, E. Levin, A. Prygarin  Nucl.Phys.A792:122-151,2007.
e-Print: arXiv:0704.2124;
E. Levin, J. Miller and A. Prygarin, 
e-Print: arXiv:0706.2944 

\bibitem{foam} A. Kovner, M. Lublinsky and U. Wiedemann, JHEP 0706:075,2007.
e-Print: arXiv:0705.1713 [hep-ph];


 \bibitem{zakopane}
A. Kovner Acta Phys.Polon.B36:3551-3592,2005.
e-Print Archive: hep-ph/0508232
\end{thebibliography}
\end{document}